\documentclass[12pt,reqno]{amsart}
\usepackage{graphicx}
\usepackage[colorlinks,linkcolor=red,anchorcolor=blue,citecolor=green]{hyperref}
\usepackage[margin=2.5cm]{geometry}

\begin{document}
\baselineskip=0.58cm 
\theoremstyle{plain}
\newtheorem{thm}{Theorem}[section]
\newtheorem{lem}{Lemma}[section]
\newtheorem{prop}{Proposition}[section]
\newtheorem{coll}{Conclusion}
\theoremstyle{remark}
\newtheorem{rem}{Remark}[section]
\title{A Predictable Rogue Wave and Generating Mechanisms}
\author{Man Jia$^{*}$ and S. Y. Lou}
\dedicatory{$^1$ Physics Department and Ningbo Collaborative Innovation Center of Nonlinear Hazard System of Ocean and Atmosphere, Ningbo University, Ningbo 315211, China}
\thanks{$^{*}$ Corresponding author: jiaman@nbu.edu.cn}

\begin{abstract}
Due to the widely applications in almost all branches of science, high dimensional KP equation is selected as universal model to describe rogue wave phenomenon. A lump is an algebraically localized wave decayed in all space directions and exists in all time. Starting from a special lump containing seven arbitrary independent parameters and four constraint conditions with all the physical properties shown, an invisible lump is found with the combination of lump part and exponential part. Because of the domination of the exponential part, the lump will be invisible in some special area, or the lump is cutoff by the induced visible soliton. While the lump part remains invariant, lump will keep its positions, path and amplitude before it is invisible. Furthermore, as a rogue wave/instanton is a localized wave decayed in all space and time directions, a rogue wave / instanton can also be produced by cutting a lump between two visible solitons. The special dispersive for the visible soliton(s) shows the soliton(s) are completely determined by the lump or the visible soliton(s) are induced by the lumps. Because the induced soliton(s) is visible, it is possible to give a prediction of the positions, the wave height and even the path for such kind of rogue waves.
\end{abstract}

\maketitle

\section{Introduction}
Rogue waves, generally accompanied by deep holes with huge wave appearance are one of the universal phenomena that have been proved in many branches of science, such as nonlinear optical systems \cite{Solli2007Optical, PhysRevLett.104.093901, PhysRevLett.103.173901}, plasmas \cite{Panwar2013Compressional, 0295-5075-96-2-25002}, fluid dynamics and atmosphere \cite{Onorato201347, M2005Rogue, Kharif2003603}, Bose-Einstein condensations (BECs) \cite{PhysRevA.80.033610}, financial systems \cite{Yan2011Financial} and microwave oscillators \cite{PhysRevLett.119.034801}, etc. Except for various physical models of the rogue wave phenomenon have been intensively developed, many laboratory experiments \cite{PhysRevLett.102.114502, rohtua, PhysRevLett.101.065303} have also been conducted to understand the physics of the giant wave character and its relation to environmental conditions. Though the origin of rogue waves is still a matter of debate \cite{Akhmediev2010}, there are two important and common features characterizing rogue wave phenomena which have been known and accepted: 1) its amplitude is more than twice (or $2.5$ times) that of the average amplitude of the significant wave height; 2) a rogue wave ``appears from nowhere and disappears without a trace'', in other words, rogue wave is believed to be unpredictable. It is obviously that rogue wave phenomena can cause dangerous problems because of its extreme wave height and unpredictability.

In the last decades, a variety of mathematical descriptions of such waves has been intensively developed. Most of the work and interpretation of the theories and experimental results are based on the nonlinear Schr\"{o}dinger equation \cite{PhysRevLett.86.5831} which can be used to describe nearly all physical phenomena in nonlinear optics, plasmas, condensed matters and fluid dynamics.

The first measurement of the rogue wave in ocean taken on the platform in $1995$ \cite{M2005Rogue} in Norway motivates to search for the physical explanation in ocean waves, especially the nonlinear shallow water waves. It is well known that the shallow water wave equation,
\begin{eqnarray}
\frac{\partial}{\partial x}\left(\frac{1}{\sqrt{gh}}\eta_t+\eta_x+\frac{3}{2h}\eta \eta_x+\frac{h^2 \gamma}{2}\eta_{xxx}\right)+\frac{1}{2}\eta_{yy}=0,\label{kp0}
\end{eqnarray}
rules the dynamics of weakly nonlinear, narrow band surface gravity waves \cite{PhysRevE.86.036305}, where $\eta=\eta(x,y,t)$ is the wave height above the constant mean height $h$, $g$ is the gravity, $\gamma=1-\tau/3$, $\tau=T/(\rho g h^2)$ is a dimensionless surface tension coefficients. The sign of $\gamma$ stands different physical meaning: when $\gamma<0$, the surface tension is large and the equation is named as KPI equation; but when it comes to $\gamma>0$, there is small surface tension with the equation being called KPII equation in soliton and integrable systems \cite{ablowitz_clarkson_1991}.

The equation (\ref{kp0}) can be rescaled into the nondimensional form
\begin{eqnarray}
u_{t}+u_{xxx}+6 u u_{xxx}+ \sigma^2 \int u_{yy} d x=0,\label{KP}
\end{eqnarray}
where $\sigma^2=\pm 1$ was firstly derived to study the evolution of long ion-acoustic waves of small amplitude propagating in plasmas under the effect of long transverse perturbations \cite{KadomtsevPetviashvili}. In this equation (\ref{KP}), $u$ denotes the wave height and $\sigma^2=\pm 1$ corresponds to the sign of $\gamma$. Later, the KP equation was widely accepted as a natural extension of the classical KdV equation to two spatial dimensions, and was derived as a model for surface and internal water waves \cite{ablowitz_segur_1979}, and in nonlinear optics \cite{PhysRevE.51.5016} and almost in all other physical fields such as in shallow water waves \cite{hammack_scheffner_segur_1989}, ion-acoustic waves in plasmas \cite{Lonngren1998}, ferromagnetics \cite{PhysRevB.77.224416}, Bose - Einstein condensation \cite{PhysRevA.77.045601} and string theory \cite{PhysRevD.36.1819}. It has been shown that the KP equation is one of the integrable equations in high dimensions in the sense of allowing a Lax pair, an infinity of conservation laws, soliton and multisoliton solutions, etc \cite{doi:10.1137/1.9781611970883}. The KP equation is also used as a classical model for developing and testing of new mathematical techniques.

In this manuscript, we select the well known high dimensional Kadomtsev-Petviashivili (KP) equation that can be found almost in all physical fields as our model to illustrate the predictable rogue wave phenomenon. The paper is organized as follows. We first establish a general form of the lump solutions for the $(2+1)$-dimensional KP equation, then extend the general form to a more general one with seven arbitrary independent parameters and four constraint conditions which is shown in Sec. II with all the physical properties of the lump being provided. By combining the lump part and exponential part, it is found in Sec. III that a lump will be invisible in some special area. Thus the lump is cutoff by the visible soliton which is totally determined by the lump. Because the lump part is invariant, the lump will keep its physical properties until it meets the induced soliton. In Sec. IV, a rogue wave/instanton is produced by cutting a lump between two solitons. Once the lump reaches a huge amplitude, it will become a rogue wave, otherwise an instanton for general amplitudes. The understanding of the physics of this kind of rogue wave phenomenon in Sec. V is very significant because the visible two solitons are induced by the lumps with the lumps becoming rogue waves/instantons which contains enough information for the lump, it is possible to give a prediction for such kind of rogue waves in some senses. The positions, the path, the wave height and even the emerge time may be predict. Sec. VI is a short summary and discussion.

\section{General Single Lump solution to KP Equation}
To explore the lump solutions to the KP equation, we rewrite the KP equation Eq.~(\ref{KP}) as
\begin{eqnarray}
(u_{t}+u_{xxx}+6 u u_{x})_x+ \sigma^2 u_{yy}=0.\label{kp1}
\end{eqnarray}
It is well known that under the transformation
\begin{eqnarray}
u=2(\ln f)_{xx}, \label{blu}
\end{eqnarray}
the KP equation Eq.~(\ref{kp1}) is transformed into the following form
\begin{eqnarray}
f f_{xt}-f_t f_x+ff_{xxxx}-4f_x f_{xxx}+3f_{xx}^2 +\sigma^2 (f_{yy}f-f_y^2)=0.\label{kpf}
\end{eqnarray}
Once the solution $f$ to the bilinear equation Eq.~(\ref{kpf}) is found, the solution to the KP equation Eq.~(\ref{kp1}) or Eq.~(\ref{kp1}) is also obtained by the transformation $u=2(\ln f)_{xx}$.

We can easily find that Eq.~(\ref{kpf}) admits a lump solution
\begin{eqnarray}
f=\sum_{i\leq j=0}^3 a_{ij} x_i x_j+f_0,\label{flump}
\end{eqnarray}
where $a_{ij}= a_{ji} = \overrightarrow{A_i} \cdot \overrightarrow{A_j}=\sum_{m=1}^M A_{im} A_{jm}$ with $\overrightarrow{A_1}=\overrightarrow{k}=|k\rangle$, $\overrightarrow{A_2}=\overrightarrow{p}=|p\rangle$, $\overrightarrow{A_3}=\overrightarrow{\omega}=|\omega\rangle$, $\overrightarrow{A_4}=\overrightarrow{\alpha}=|\alpha\rangle$ are vectors related to real constants $k_m$, $p_m$, $\omega_m$ and $\alpha_m$, $m=1,\ 2,\ \cdots, M$, and $x_1=x$, $x_2=y$, $x_3=t$, $x_0=1$ with $\omega_m$ and $f_0$ being
\begin{eqnarray}
\omega_m=\frac{\sigma^2(a_{22} k_m-2 a_{12} p_m)}{a_{11}},\label{omegam}
\end{eqnarray}
\begin{eqnarray}
f_0=-a_{00}+\frac{a_{01}^2a_{22}-2 a_{01} a_{02} a_{12}+a_{02}^2a_{11}}{a_{11}a_{22}-a_{12}^2}-\frac{3 \sigma^2 a_{11}^3}{a_{11}a_{22}-a_{12}^2}.\label{f00}
\end{eqnarray}

For example, by taking $M=3$, though the vectors $\overrightarrow{A_i}$ contains twelve parameters for $k_m$, $p_m$, $\omega_m$ and $\alpha_m$, $m=1,\ 2,\ 3$, there are eleven independent parameters indeed according to the expansion expression of $f$ in Eq.~(\ref{flump}) with the four constraint conditions $\omega_m$, $m=1,\ 2,\ 3$ and $f_0$ are provided by
\begin{eqnarray}
\omega_1=-\frac{\sigma^2[+k_1(p_1^2-p_2^2-p_3^2)+2p_1(k_2 p_2+k_3 p_3)]}{k_1^2+k_2^2+k_3^2},\label{omega31}
\end{eqnarray}
\begin{eqnarray}
\omega_2=-\frac{\sigma^2[k_2(-p_1^2+p_2^2-p_3^2)+2p_2(k_1 p_1+k_3 p_3)]}{k_1^2+k_2^2+k_3^2},\label{omega32}
\end{eqnarray}
\begin{eqnarray}
\omega_3=-\frac{\sigma^2[k_3(-p_1^2-p_2^2+p_3^2)+2p_3(k_1 p_1+k_2 p_2)]}{k_1^2+k_2^2+k_3^2},\label{omega33}
\end{eqnarray}
\begin{eqnarray}
&& f_0=-[(k_1 p_2-k_2p_1)^2 +(k_2 p_3-k_3 p_2)^2+(k_1 p_3-k_3 p_1)^2]^{-1} \nonumber \\ && \qquad \times \{[\alpha_1(k_2p_3-k_3p_2)+\alpha_2(k_3 p_1 -k_1 p_3)+\alpha_3(k_1 p_2 \nonumber \\ && \qquad -k_2 p_1)]^2+3 \sigma^2 (k_1^2 +k_2^2+k_3^2)^3\}.\label{f30}
\end{eqnarray}
That means we have eleven independent parameters with seven arbitrary constants and four constraint conditions in the solution. Compared with \cite{Ma20151975} possessing six free arbitrary parameters and three constraint conditions and other known results \cite{Wang2017Some} and \cite{Manakov1977205, doi:10.1063/1.523550}, the result presented here is more general.

Then the solution $f$ to the bilinear KP equation Eq.~(\ref{kpf}) in expansion form is
\begin{eqnarray}
&& f_{lump}\equiv a_{11} x^2+2 a_{12} x y+2 a_{13} x t+a_{22}y^2+ 2 a_{23} y t+a_{33}t^2+2 a_{01} x+2 a_{02} y\nonumber \\ && \qquad+2 a_{03} t+\frac{a_{01}^2a_{22}-2 a_{01} a_{02} a_{12}+ a_{02}^2 a_{11}} {a_{11}a_{22} -a_{12}^2}-\frac{3 \sigma^2 a_{11}^3}{a_{11}a_{22}-a_{12}^2},\label{flump3}
\end{eqnarray}
with the corresponding more general single lump solution to KP equation Eq.~(\ref{kp1}) being
\begin{eqnarray}
u_{lump}=\frac{4 a_{11}}{f_{lump}}-\frac{8(a_{11}x+a_{12}y+a_{13}t+a_{01})^2}{f_{lump}^2}.\label{ulump3}
\end{eqnarray}

According to the transformation $u=2(\ln f)_{xx}$, we have to put a constraint
\begin{eqnarray}
f>0,\nonumber
\end{eqnarray}
which leads to the requirement of $f_0>0$ to insure $u$ analytical. From the special expressions of $f_0$ given by Eq.~(\ref{f30}) for $n=3$, we know $\sigma^2$ have to be fixed as $-1$.

To see the physical properties of a lump more concretely, we present the positions of the lump described by $[x,\ y]$ at any time. Generally, the positions of a lump can be provided by using a curve related to time $t$. By some simple differential calculations of $u_x=u_y=0$, the positions of a lump is proved to be
\begin{eqnarray}
x=-\frac{\sigma^2 a_{22} t}{a_{11}}-\frac{a_{01}a_{22}-a_{12}a_{02}}{a_{11} a_{22}-a_{12}^2},\qquad y=\frac{2 \sigma^2 a_{12} t}{a_{11}}+\frac{a_{01}a_{12}-a_{11}a_{02}}{a_{11} a_{22}-a_{12}^2},\label{top}
\end{eqnarray}
or the lump moves along the straight line
\begin{eqnarray}
y=-\frac{2 a_{12} x}{a_{22}}-\frac{a_{01}a_{12}a_{22}+a_{11}a_{02}a_{22}-2a_{12}^2 a_{02}}{a_{22}(a_{11} a_{22}-a_{12}^2)}.\label{path}
\end{eqnarray}

\begin{figure}
\centering\includegraphics[width=5cm]{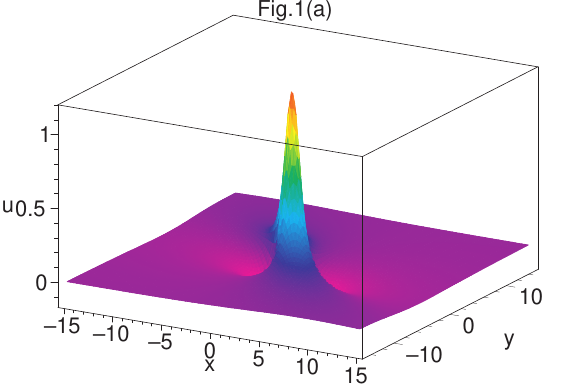}
\centering\includegraphics[width=5cm]{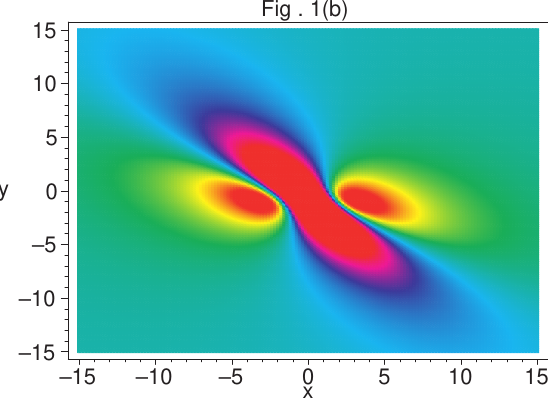}
\centering\includegraphics[width=5cm]{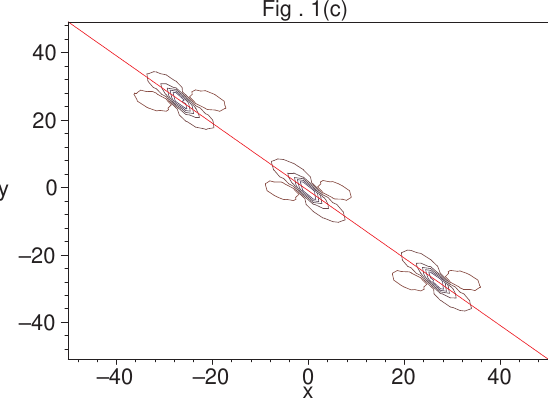}
\caption{\label{fig1} The exhibition of the lump with the parameters selected as $k_1=1$, $k_2=1$, $k_3=-\frac{1}{2}$, $p_1=1$, $p_2=1$, $p_3=1$, $\alpha_2=1$ and $\alpha_3=1$. (a) the structure at $t=0$; (b) the corresponding density plot; and (c) the positions of the lump at $t=-20$, $t=0$ and $t=20$, respectively.}
\end{figure}

The amplitude of a lump is also considered which is defined by the difference between the minimum and maximum values a lump could have. The positions of the minimum and maximum value of a lump can be obtained by solving the system $\{u_x=0,\ u_y=0\}$. It is not difficult to verify that the amplitude of a lump is
\begin{eqnarray}
A_{lump}=\left |-\frac{3\sigma^2 (a_{11} a_{22}-a_{12}^2)}{2 a_{11}^2}\right | ,\label{amp}
\end{eqnarray}
which reveals that the amplitude of a lump is also a constant related to the arbitrary constants of $a_{11}$ and $a_{22}$. A special lump Eq.~(\ref{ulump3}) is depicted in FIG.~\ref{fig1} with its path.

\section{An Invisible Lump}
A lump will be invisible in some special area. For example, if we take
\begin{eqnarray}
f_{lumpoff} \equiv f_{lump}+a_0 \textrm{e}^{k_0 x+p_0 y+ \omega_0 t+x_0},\label{rflumpoff}
\end{eqnarray}
which consists of lump part and exponential part, the lump part will become invisible at the special area of $k_0x+p_0 y+\omega_0 t+x_0>0$ because of the domination of the exponentiation part. That is to say, the lump only exists at the special area of $k_0x+p_0 y+\omega_0 t+x_0<0$.

Substituting Eq.~(\ref{rflumpoff}) into the KP equation Eq.~(\ref{kpf}) with a direct calculation, we find $k_0$, $p_0$ and $\omega_0$ are
\begin{equation}
k_0^2=-\frac{\sigma^2(a_{11} a_{22}-a_{12}^2)}{3a_{11}^2},\label{k01}
\end{equation}
\begin{equation}
p_0=\frac{k_0 a_{12}}{a_{11}}, \label{p01}
\end{equation}
\begin{equation}
\omega_0=-\frac{k_0^4+\sigma^2 p_0^2}{k_0},\label{o01}
\end{equation}
where $a_{ij}$ is related to arbitrary constants of $k_m$, $p_m$ and $\alpha_m$ with $a_0$ and $x_0$ being arbitrary constants.

The result is quite interesting because it demonstrates a special soliton induced by lump. Eq.~(\ref{o01}), considered as a special dispersive relation, indicates $\omega_0$ is related to $k_0$ and $p_0$, while $k_0$ and $p_0$ are completely determined by the lump according to Eqs.~(\ref{k01}) - (\ref{p01}). Thus the soliton (the exponential part) is induced by the lump. The existence of the soliton is based on the existence of the lump. If the lump does not exist, the soliton will also disappear. Once the soliton is induced, dut to the dominantion of the exponentiation part, the lump will be invisible. The lump becomes a lumpoff which is cutoff by the soliton induced by itself.

According to Eqs. (\ref{k01}) - (\ref{o01}), the parameters of the induced soliton may be chosen as $k_0>0$ or $k_0<0$. If $k_0>0$, the lump is visible at the area of $k_0x+p_0 y+\omega_0 t+x_0<0$, which is cutoff by the induced soliton and eventually disappears. Otherwise, a revers process for $k_0<0$ showing the interaction between the invisible lump and the induced soliton with the lump emerging at $k_0x+p_0 y+\omega_0 t+x_0>0$.

Because we take the combination of unchanged lump part and exponential part, it is easy to find the corresponding positions of the lump are still determined by Eq.~(\ref{top}).

\begin{figure}
\includegraphics[width=0.4\textwidth]{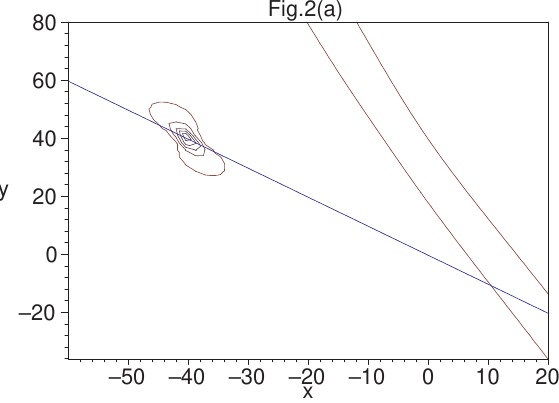}
\includegraphics[width=0.4\textwidth]{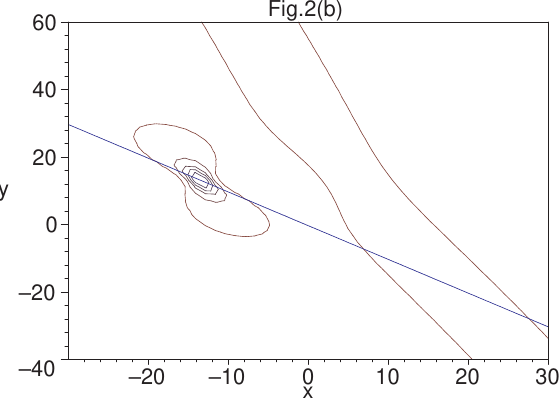}
\includegraphics[width=0.4\textwidth]{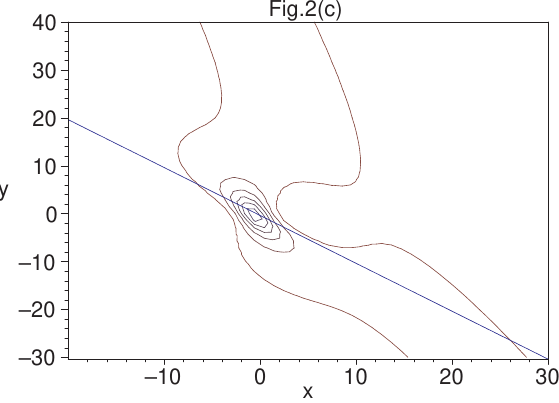}
\includegraphics[width=0.4\textwidth]{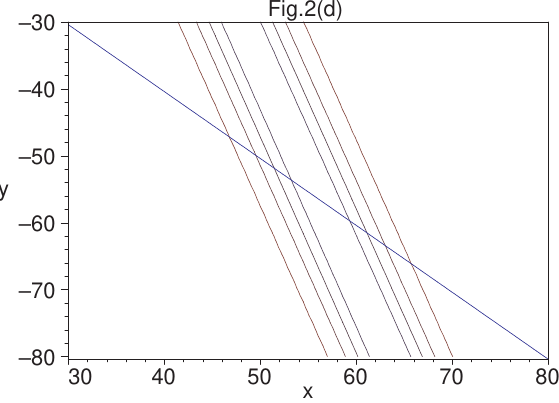}
\caption{\label{fig3} The plot of a lumpoff with parameters selected in Eq.~(\ref{lumpoff1}) where the blue line is the path denoting the positions of the lump.}
\end{figure}

FIG. \ref{fig3} is the contour plot of the lumpoff where the blue line is the path showing the positions of the lump according to Eq.~(\ref{top}) at (a) $t=-60$, (b) $t=-20$, (c) $t=0$ and (d) $t=250$, respectively. It can be found that the lump remains its path until it is cutoff by the induced soliton. The parameters are selected as
\begin{eqnarray}
&&k_1=1, \quad k_2=\frac{1}{2}, \quad k_3=1, \quad p_1=1, \quad p_2=\frac{1}{2}, \quad p_3=-\frac{1}{2},\nonumber \\ && \alpha_1=0, \quad \alpha_2=1, \quad \alpha_3=0, \quad x_0=1, \quad a_0=3,\label{lumpoff1}
\end{eqnarray}
for $\{+k_0,\ +p_0,\ -\omega_0\}$.

\begin{figure}
\includegraphics[width=0.4\textwidth]{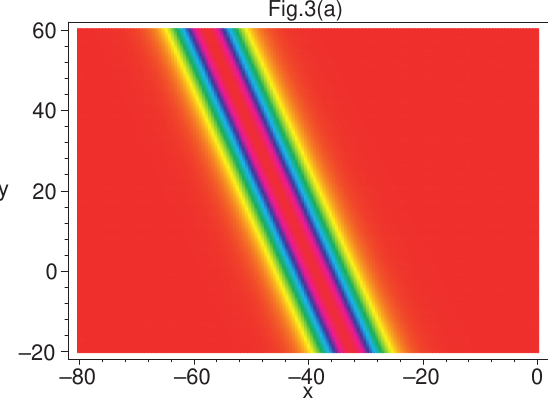}
\includegraphics[width=0.4\textwidth]{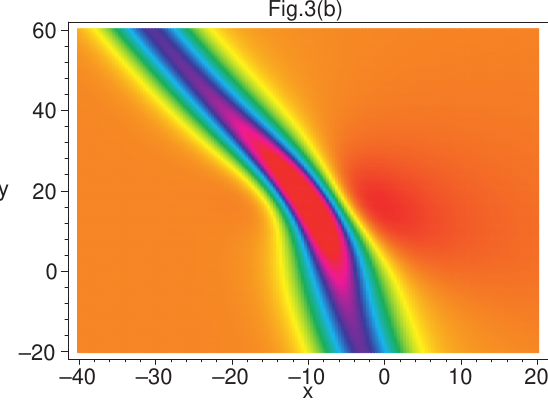}
\includegraphics[width=0.4\textwidth]{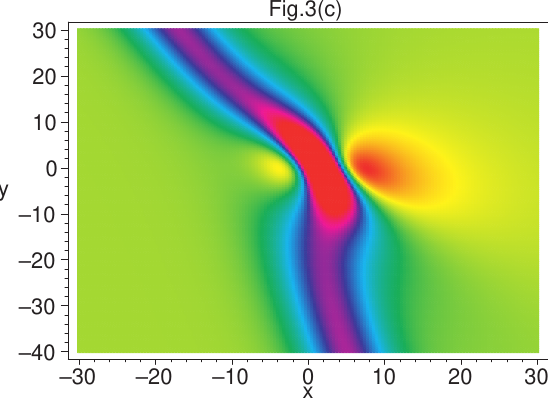}
\includegraphics[width=0.4\textwidth]{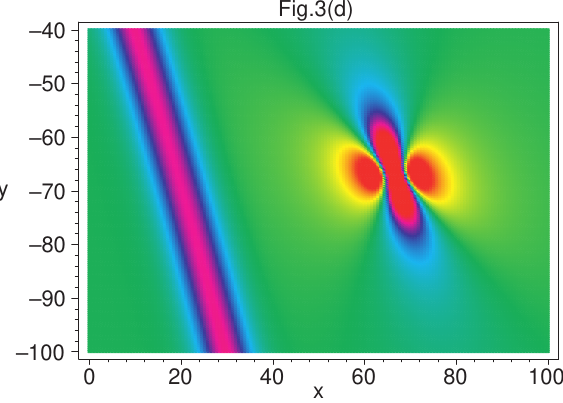}
\caption{\label{fig5} The density plot of the lumpoff with $k_0<0$ at (a) $t=-250$, (b) $t=-25$, (c) $t=0$ and (d) $t=50$, respectively.}
\end{figure}

FIG.~\ref{fig5} is a density plot with the same parameters of Eq.~(\ref{lumpoff1}) for $\{-k_0,\ -p_0,\ +\omega_0\}$. The lump is first invisible and ultimately separates out from the induced soliton.

\section{Rogue Wave/Instanton}

A rogue wave/instanton is a localized wave decayed in all space and time directions. In fact, in quantum field theories the instantons studies allow scientists to see the previously hidden logarithmic structure of the states and operators \cite{frenkel_losev_nekrasov_2011, Frenkel2007215}. Physicists believe that instantons are the key to explore the interactions principle in the standard model. Studies exhibits that instantons have been shown in integrable systems, such as DS equation \cite{1402-4896-65-1-001} and many $(2+1)$-dimensional models \cite{PhysRevE.66.046601} via the multiple linear variable separation approach.

The rogue wave/instanton solutions to the KP equation is construct by
\begin{eqnarray}
f_{instanton}\equiv f_{lump}+a_1 \cosh (k_0 x+p_0 y+ \omega_0 t+x_0)+g_0, \label{finstanton}
\end{eqnarray}
where $f_{lump}$ is shown in Eq.~~(\ref{flump3}), and $k_0$, $p_0$, $\omega_0$ satisfy the constraint conditions of Eqs.~~(\ref{k01}) - (\ref{o01}), with $a_1$ and $g_0$ being two arbitrary constants to be determined. As the visible lump being confined in a special area for $k_0 x+p_0y+\omega_0 t+x_0<0$, the lump should be confined in a twin-soliton induced by itself. In other words, the lump's emerge for the area $k_0x+p_0 y+\omega_0 t+x_0 \sim 0$ for a special time leads to the lump becomes a rogue wave/instanton.

Substituting Eq.~~(\ref{finstanton}) into the bilinear form of the KP equation Eq.~~(\ref{kpf}) and eliminating all the coefficients of the polynomials of \{$x$,\ $y$,\ $t$,\ $\cosh$\ and/or $\sinh$\} by using the known results and constraints, one can easily find the only possible choice for $g_0$ is
\begin{equation}
g_0=-\frac{\sigma^2 a_1^2 (a_{11} a_{22}-a_{12}^2)}{12a_{11}^3},\label{g0}
\end{equation}
with $a_1$ being an arbitrary constant.

The rogue wave/instanton is obtained by cutting the lump between two solitons with the lump being visible in the area $k_0x+p_0 y+\omega_0 t+x_0 \sim 0$ for a special time. Due to the existence of the lump, a twin-soliton is induced according to the special dispersion relation Eq.~(\ref{o01}) which is visible all time because of the domination of the $\cosh$ part. The visible of solitons leads to the invisible of the lump, thus the lump is visible only when it moves to the line $k_0 x+p_0 y +\omega_0 t+x_0\sim 0$. Once the lump reaches a large amplitude, it will become a rogue wave, or be an instanton for general amplitudes.

Because of the unchanged lump part in solution Eq.~(\ref{finstanton}), the invisible lump will remains its path and positions according to Eq.~(\ref{top}) or Eq.~(\ref{path}). Once the invisible lump comes to $k_0 x+p_0 y +\omega_0 t+x_0\sim 0$, it will appear until it reaches peak at the cross point of the centerline $k_0 x+p_0 y+\omega_0 t+x_0=0$ of the two solitons. That means the rogue wave will appear or reach its peak at
\begin{eqnarray}
t=-\frac{9k_0a_{11}^3(k_0a_{10}-a_{11}x_0)}{2(a_{11}a_{22}-a_{12}^2)^2},\label{et}
\end{eqnarray}
with the place
\begin{eqnarray}
&&x=-\frac{a_{10}a_{22}-a_{12}a_{20}}{a_{11} a_{22}-a_{12}^2}-\frac{9 k_0 a_{22} a_{11}^2(k_0 a_{10}-a_{11}x_0)}{\sigma^2(a_{11} a_{22}-a_{12}^2)^2},\nonumber \\  && y=\frac{a_{01}a_{12}-a_{11}a_{02}}{a_{11} a_{22}-a_{12}^2}\frac{9k_0 a_{12}a_{11}^2(k_0 a_{10}-a_{11}x_0)}{\sigma^2(a_{11} a_{22}-a_{12}^2)^2}.\label{exy}
\end{eqnarray}

Moreover, the maximum value of wave hight of the instanton/rogue wave is
\begin{eqnarray}
A_{instanton}=\left |\frac{8a_{11}(a_{11} a_{22}-a_{12}^2)}{a_1 (a_{11} a_{22}-a_{12}^2)+6a_{11}^3}\right|,\label{ainst}
\end{eqnarray}
which indicates that the amplitude is related to the soliton's parameter $a_1$ and the lump part of $a_{11}$, $a_{12}$, $a_{22}$ by calculating the value of $u$ when the lump arrive at the center of the twin-soliton.

\begin{figure}
\includegraphics[width=0.4\textwidth]{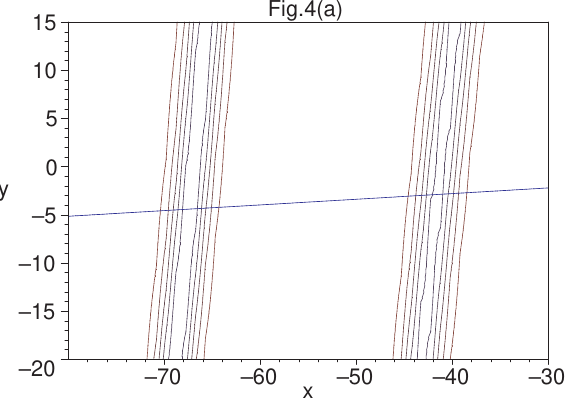}
\includegraphics[width=0.4\textwidth]{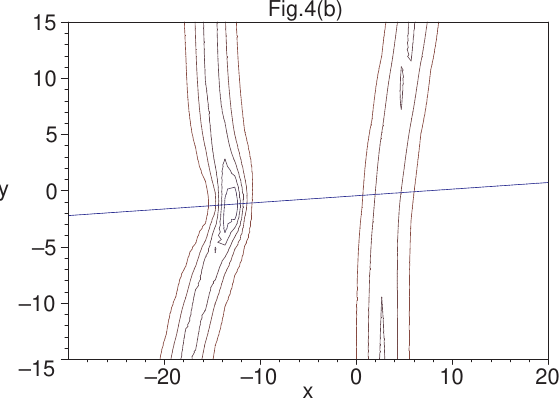}
\includegraphics[width=0.4\textwidth]{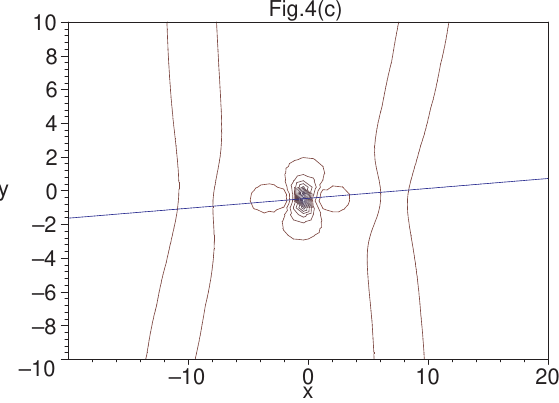}
\includegraphics[width=0.4\textwidth]{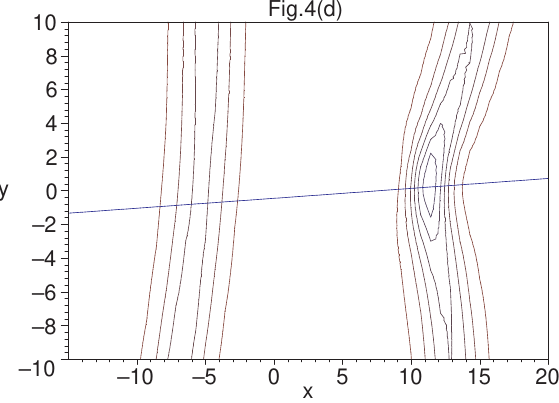}
\includegraphics[width=0.4\textwidth]{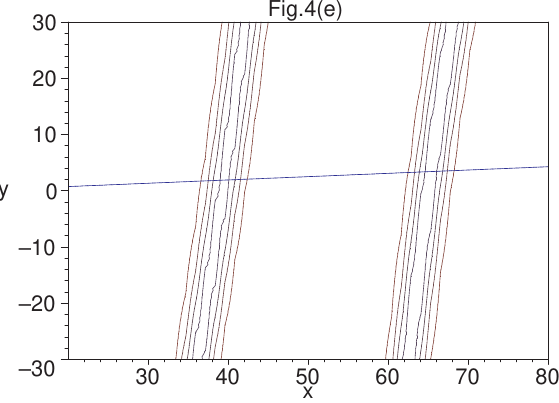}
\includegraphics[width=0.4\textwidth]{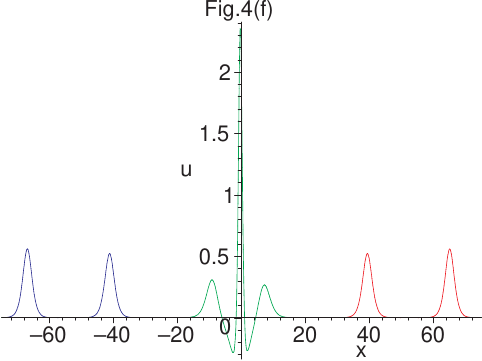}
\caption{\label{fig6} The plot of the rogue wave with parameters selections of Eq.~ (\ref{ccinst1}) with (f) the corresponding wave height in $y=0$ plane at $t=-50$ in blue, $t=0$ in green and $t=50$ in red. }
\end{figure}

The contour plot of a rogue wave with its path is exhibited in FIG.~\ref{fig6} at (a) $t=-50$, (b) $t=-5$, (c) $t=0$, (d) $t=5$ and (e) $t=50$, respectively. And (f) is the change of wave height in the plane of $y=0$  $t=-50$ in blue, $t=0$ in green and $t=50$ in red where the parameters are selected as
\begin{eqnarray}
&&k_1=\frac{1}{4}, \quad k_2=1, \quad k_3=\frac{1}{2}, \quad p_1=\frac{3}{2}, \quad p_2=-1,\quad p_3=1,\nonumber \\ && \alpha_1=1, \quad \alpha_2=0, \quad \alpha_3=\frac{1}{2}, \quad x_0=1,\quad a_1=\frac{1}{20}, \quad \sigma^2=-1. \label{ccinst1}
\end{eqnarray}
It is easy to find the rogue wave will reach its peak at about $t=-0.27$ in $\{x=-1.30, y=-0.53\}$ with the amplitude being $A = \frac{598080}{141763}$ according to Eqs.~(\ref{et}) - (\ref{ainst}).

\begin{figure}
\includegraphics[width=0.4\textwidth]{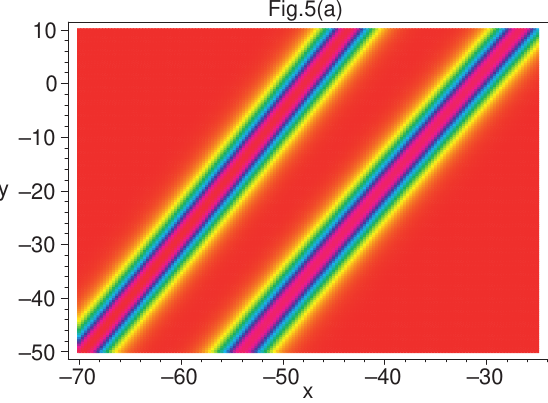}
\includegraphics[width=0.4\textwidth]{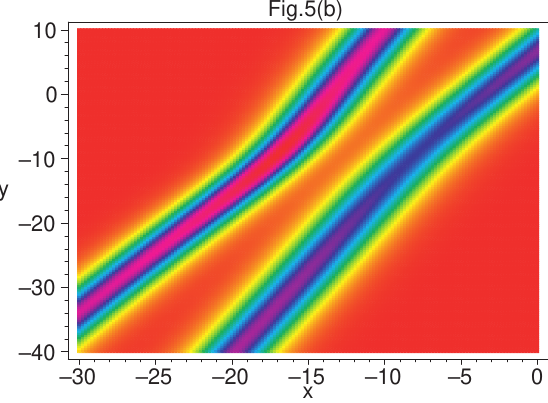}
\includegraphics[width=0.4\textwidth]{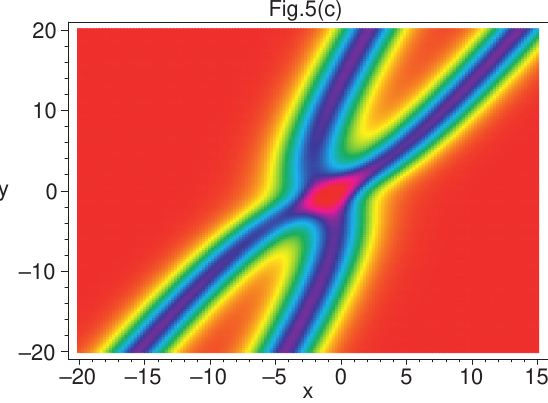}
\includegraphics[width=0.4\textwidth]{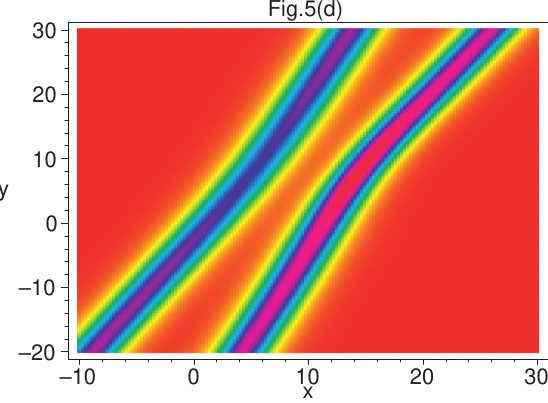}
\includegraphics[width=0.4\textwidth]{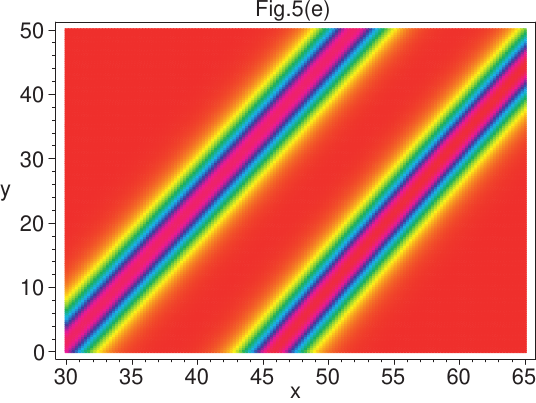}
\includegraphics[width=0.4\textwidth]{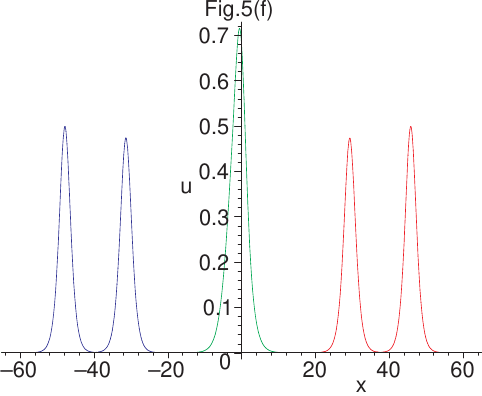}
\caption{\label{fig8} The instanton solution of the KP equation given by Eq.~(\ref{ccinst3}) with (f) being the wave shape in $y=0$ plane.}
\end{figure}

Except for rogue wave, it is possible to find an instanton with small amplitude under suitable parameters selections from Eq.~(\ref{ainst}). If the arbitrary parameters are selected as
\begin{eqnarray}
&&k_1=\frac{1}{4}, \quad k_2=1, \quad k_3=\frac{1}{4}, \quad p_1=\frac{3}{2},\quad p_2=-1,\quad p_3=\frac{1}{2}, \nonumber \\ && \alpha_1=1, \quad \alpha_2=0, \quad \alpha_3=\frac{1}{2}, \quad x_0=1,\quad a_1=10, \quad \sigma^2=-1, \label{ccinst3}
\end{eqnarray}
we have an instanton with the amplitude being only about $A=\frac{8496}{11627}$ which can be verified by substituting the parameters into the amplitude expression. FIG.~(\ref{fig8}) exhibits the instanton at (a) $t=-50$, (b) $t=-10$, (c) $t=0$, (d) $t=10$ and (e) $t=50$, and (f) is the wave shape in the plane of $y=0$ at $t=-50$ in blue, $t=0$ in green and $t=50$ in red.

\section{Generating Mechanisms for a Predictable Rogue Wave}
Our results shows novel generating and prediction mechanism for a type of rogue waves.

Lump is localized wave decayed in all space directions and exists in all time, but it will be invisible in a special area due to the domination of the exponential function. It is interesting that a special dispersive relation for the soliton(s) is found showing that the soliton(s) is totally determined by the lump. We should emphasize that the soliton(s) is induced by the lump. If there is no lump, there is no soliton(s) with the special dispersion relation. Furthermore, whence soliton(s) is induced by the lump, the lump will be cutoff by the induced soliton(s) and become invisible. When a soliton is induced by the lump, the lump will be cutoff and become invisible to a lumpoff. When two solitons are induced by the lump, the lump will become a rogue wave (or instanton) and can only be visible at an instant time.

Because the two solitons includes enough information ($k_0$, $p_0$ and $\omega_0$) of the invisible lump (algebraic) part ($a_{11}$, $a_{22}$ and $a_{12}$), The position, the path and the wave height, even the emerge time of the rogue wave may be predict.

It should be mentioned that this kind of rogue wave is found under the selection of $\sigma^2=-1$ which means the KP equation possesses the lump solution under large surface tension. Though the lump has not yet been observed experimentally, our results provides a theoretical way to understand rogue waves and its generating mechanics.

\section{Summary and Discussion}
In summary, we choose the KP equation as our universal model to present rogue wave phenomenon not only because it is widely used in almost all branches of science, but also it is high dimensional model. The widely applications and generality of KP equation makes it possible to explain almost all the rogue wave phenomenon. The progress in the understanding of the physics of the rogue wave phenomenon and development of adequate mathematical models is very significant.

Though lump solutions to KP equation have been presented by many authors, we find a special lump containing seven independent parameters and four constraint conditions which is believed to be a more general form. The positions, the path and the amplitude of the lump are also provided.

By using the combination of lump part and exponential function, lumpoff and instanton/rogue wave solutions to the KP equation are successfully illustrated. A lump will become lumpoff or instanton/rogue wave when it is cutoff by induced soliton(s). If there is no lump, there is no soliton(s) with the special dispersion relation. Because the soliton(s) includes enough information ($k_0$, $p_0$ and $\omega_0$) of the invisible lump, it provide a possible way to predict this kind of rogue wave.

\section*{Acknowledgments}
The authors are indebt to thank Prof B. F. Feng for their helpful discussions. The work is supported by NNSFC (Nos. 11675084 and 11975131), and the authors are sponsored by K. C. Wong Magna Fund in Ningbo University.

\providecommand{\newblock}{}

\end{document}